\documentclass[preprint,showpacs,aps]{revtex4}

\usepackage{epsfig}

\begin{document}

\title{\Large \bf Computer simulations of two-dimensional melting with
dipole-dipole interactions}

\author{\bf S.Z. Lin$^{1}$, B. Zheng$^{1,2}$ and S. Trimper$^{2}$}

\affiliation{$^1$ Zhejiang University, Zhejiang Institute of Modern
Physics, Hangzhou 310027, P.R. China\\
$^2$ FB Physik, Universit\"at -- Halle, 06099 Halle, Germany\\}

\begin{abstract}
We perform molecular dynamics and Monte Carlo simulations of
two-dimensional melting with dipole-dipole interactions. Both
static and dynamic behaviors are examined. In the isotropic liquid
phase, the bond orientational correlation length $\xi_{6}$ and
susceptibility $\chi_{6}$ are measured, and the data are fitted to the
theoretical ansatz. An algebraic decay is detected for both
spatial and temporal bond orientational correlation functions in
an intermediate temperature regime, and it provides an explicit
evidence for the existence of the hexatic phase. From the
finite-size scaling analysis of the global bond orientational
order parameter, the disclination unbinding temperature $T_{i}$ is
estimated. In addition, from dynamic Monte Carlo simulations of
the positional order parameter, we extract the critical exponents
at the dislocation unbinding temperature $T_{m}$. All the results
are in agreement with those from experiments and support the KTHNY
theory.
\end{abstract}


\maketitle

\maketitle

\section{Introduction}

Two-dimensional melting has been intensively studied in the past
years, but it is still not completely understood
\cite{str88,gla93,das99}. Melting in two dimensions is quite
different from its counterpart in three dimensions, for a true
long-range positional order doesn't exist in two-dimensional
systems. The absence of a conventional long-range order at
non-zero temperature was first pointed out by Mermin and Wagner
\cite{mer66}. Nevertheless, another long-range order, which is
called the bond orientational order, can be observed in the solid
phase \cite{mer68}.

There exist several possible theoretical descriptions of melting in
two-dimensional systems. The KTHNY theory, developed by Halperin,
Nelson and Young \cite{hal78,nel79,you79}, predicts that a third
phase, the so-called hexatic phase, may exist between solid and
liquid states in a portion of the phase diagram. The system first melts
from the solid state to the hexatic state due to the unbinding of
dislocation at a temperature $T_{m}$, and then melts from the hexatic
state to the liquid state at the disclination
unbinding temperature $T_{i}$. Both transitions are
Kosterlitz-Thouless phase transitions \cite{kos73}. Naturally, the
KTHNY theory only describes a possible scenario. It is also
possible that anyone or both of the continuous transitions are of
first order, and even that there is a direct first-order
transition from the solid state to the isotropic liquid state.

Even though quite some experiments supported the KTHNY
theory \cite{dav84,gal82,guo83,mct82,nie87,gre87,kus94,zah99}, most
early works of computer simulations on two-dimensional melting
favored a first-order phase transition, and the {\it hexatic} phase was
not observed. For example, for the systems with dipole-dipole
interactions, Kalia and Vashishta \cite{kal81} observed a
superheating and supercooling, as well as a latent heat in
two-dimensional melting, and concluded that the phase transition
is of first order. Later, Bedanov, Gadiyak and Lozovik
\cite{bed82} found that both the positional and bond
orientational order vanished simultaneously at the melting point,
and the hexatic phase didn't exist. Similar results have been found for
other systems \cite{tox81,str84,bra01,web95}. Even for the
simplest system, the hard disk model, there was no consensus about
the melting mechanism \cite{zol89,zol92,web95,fer95,fer97}.

In 1996, Bagchi {\it et al.} \cite{bag96} carried out a
finite-size scaling analysis of the bond orientational order
parameter in a system interacting via a repulsive $1/r^{12}$
potential, and found that the results were in agreement with the
KTHNY theory, even though no conclusive evidence for the hexatic
phase was observed. Later, extensive Monte Carlo simulations of
the hard disk model were performed by Jaster \cite{jas98,jas99}.
Numerical behaviors of the susceptibility, spatial bond
orientational correlation length and pressure, support the
KTHNY theory. But the algebraic decay of the bond
orientational correlation function was still
not shown \cite{jas04}. Recently, Monte Carlo simulations
of a two-dimensional electron system with a $1/r$ interacting
potential have been performed by He {\it et al.} \cite{he03}. An
algebraic decay of the bond orientational correlation function is
observed, and it explicitly reveals the existence of the hexatic phase.
In principle, however, the finite-size effect and coexistence of
liquid and solid phases may also lead to such an algebraic decay.
One needs to carefully rule out these possibilities.
On the other hand, in all these numerical simulations
of the bond orientational order,
the static behavior of the melting is mainly concerned, and the
dynamics is not touched so much.

Recently, more progress in computer simulations has been made
in understanding two-dimensional melting, for example,
on the roles of the polydispersity \cite{pro04,wat05}
and external fields \cite{str02}, and on the structural change
during the melting \cite{mou05}.
Especially, some experiments show much interest in a two-dimensional
system with dipole-dipole interactions \cite{zah99,zah00}.
The algebraic decay of the spatial and temporal correlation functions
are observed and the {\it dynamic behavior} is found to be very
relevant for two-dimensional melting. From the view of numerical
simulations, the two-dimensional system with dipole-dipole
interactions is not much understood. The purpose of this article
is to present systematic computer simulations of two-dimensional
melting with dipole-dipole interactions. Main results are obtained
with molecular dynamics simulations, and Monte Carlo simulations
are also performed in some cases and for confirmation. Both static
and dynamic behavior will be examined, and an emphasis is given to
the algebraic decay of both the {\it spatial} and {\it temporal}
bond orientational correlation functions in the hexatic phase.

The article is organized as follows. In Sec. II, the model and
numerical methods will be described. In Sec. III, numerical
results will be presented for both static and dynamic behavior.
Finally it comes the conclusion.

\section{Model and method}

\subsection{The model}

In this article, we consider a two-dimensional
dipolar system whose Hamiltonian can be written as
\begin{equation}\label{ur}
H=\sum_{i}^{N}\frac{p_{i}^{2}}{\mu_{i}}+\frac{1}{2}\sum_{i=1}^{N}\sum_{j\neq i}^{N}\{ \frac{\overrightarrow{m_{i}}\cdot \overrightarrow{m_{j}}}{|\overrightarrow{r_{ij}}|^{3}}
-3\frac{(\overrightarrow{m_{i}}\cdot \overrightarrow{r_{ij}})(\overrightarrow{m_{j}}\cdot \overrightarrow{r_{ij}})}{|\overrightarrow{r_{ij}}|^{5}}
  \},
\end{equation}
where $p_{i}, \mu_{i}$ and $\overrightarrow{m_{i}}$ are the
momentum, mass and magnetization of the \emph{i}th dipole
respectively, and $N$ is the total number of particles.
In order to mimic the experiments in Refs.
\cite{zah99,zah00} and to simplify the problem, we assume the
dipoles are aligned perpendicular to the surface.
Thus, Eq. (\ref{ur}) can be reduced to
\begin{equation}\label{urr}
H=\sum_{i}^{N}\frac{p_{i}^{2}}{\mu_{i}}+\frac{1}{2}\sum_{i=1}^{N}\sum_{j\neq
i}^{N}\frac{m_{i}m_{j}}{|\overrightarrow{r_{ij}}|^{3}} .
\end{equation}
For convenience in numerical simulations, we rewrite Eq. (\ref{urr}) as
\begin{equation}\label{urrr}
H=\sum_{i}^{N}\frac{p_{i}^{2}}{\mu}+\frac{1}{2}\sum_{i=1}^{N}\sum_{j\neq
i}^{N}\epsilon(\frac{\sigma}{r_{ij}})^{3} ,
\end{equation}
where we have assumed the mass and magnitude of the magnetization of the dipoles are identical. For simplification, the reduced units are adopted, in
which the parameters $\epsilon$ and $\sigma$, Boltzmann
constant $k_{B}$, and mass $\mu$ of the dipoles are set to $1$. The thermodynamic observables are determined only by a dimensionless
constant $\Gamma=\epsilon \sigma^{3}(\pi n)^{3/2}/kT$ \cite{hag05} , where $n=N/V$ is the $2D$ volume fraction of the dipoles.

The reasons we choose this model are: (i) this model lacks extensive numerical study, and the existing results do not favor the KTHNY theory; (ii) there are unambiguous
experimental results of such system \cite{zah99,zah00},
to which we may compare our results.

\subsection{Numerical methods}

\begin{figure}[h]
\epsfysize=6.0cm \epsfclipoff \fboxsep=0pt
\setlength{\unitlength}{1cm}
\begin{picture}(10,7.0)(0,0)
\put(-4.3,0.5){\epsffile{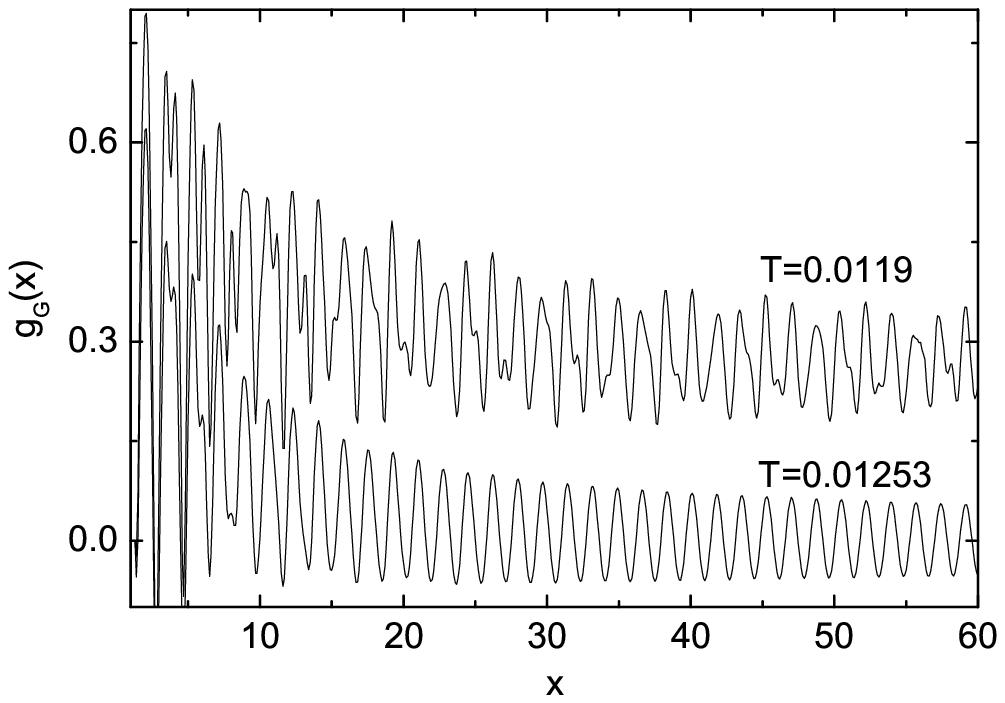}} \epsfysize=6.0cm
\put(4.5,0.5){\epsffile{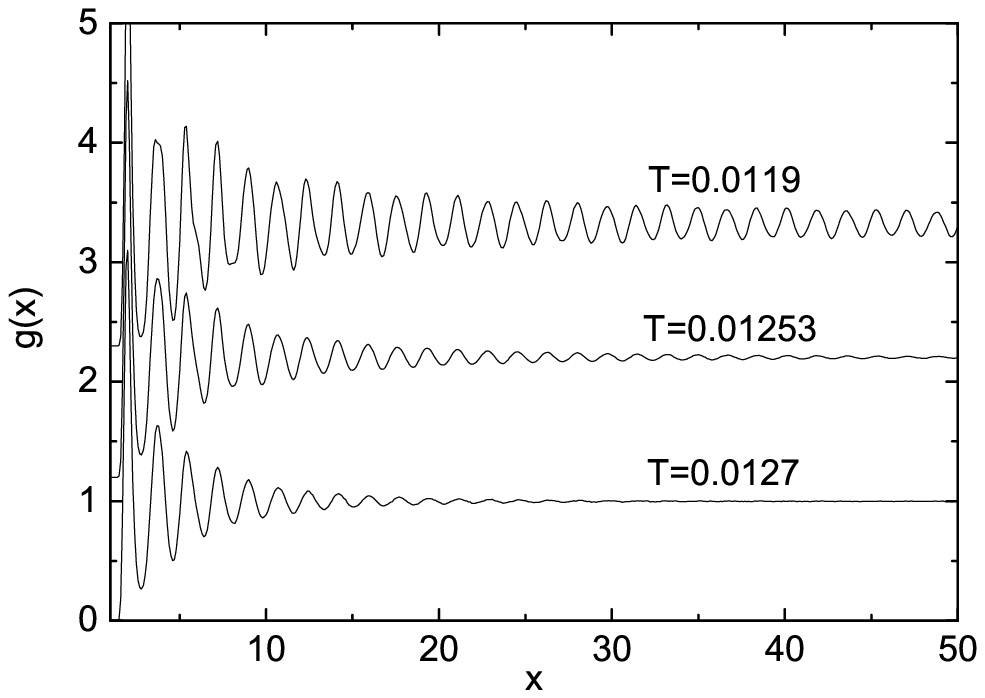}} \put(-2.5,0.2){(a)}
\put(6.2,0.2){(b)}
\end{picture}
\caption{(a) The positional correlation function $g_{G}(x)$ at
$T=0.01253$ in the hexatic phase (lower curve) and $0.0119$ in the
solid phase (upper curve) plotted vs. $x$ on a linear scale. (b)
The pair distribution function $g(x)$ plotted vs. $x$ on a linear
scale. The upper two curve are shifted upward for clarity. The curves
at $T=0.0119$, $T=0.01253$ and $T=0.0127$ show features in the
solid, hexatic and liquid phases.}\label{f4}
\end{figure}

In our simulations, particles are put in a rectangular box with a
size ratio $2:\sqrt{3}$, {\it the density of the particles is fixed
to be $1/(2\sqrt{3})$}, and the number of the particles is taken to
be from $1024$ to $32768$. The linear size $L$ of the system is
related to the total number $N$ of particles by $L=2\sqrt N$.
Periodic boundary conditions are used in simulations, and the
dipole-dipole potential is truncated at $10$. In two dimensions,
such a truncation is reasonable. Actually, the correction of the
potential to the truncation is
$u_{c}=\int_{0}^{2\pi}\int_{r_{c}}^{+\infty}g(r)\epsilon(\frac{\sigma}{r})^{3}rdrd\theta$;
assuming $g(r)=1$, it leads to $u_{c}=2\pi\epsilon\sigma^{3}/r_{c}$,
which decays to zero with $r_c$. In fact, the main parts of the
simulations are carried out in the hexatic and liquid phases where
$g(r)$ quickly stabilizes at a constant which is smaller than $1$
(see Fig.~\ref{f4} (b)).

In order to confirm the truncating procedure, we have performed the
simulations at different truncating distances, and find that the
difference is negligibly small. In addition, extra simulations using
the Ewald Summation technique\cite{ewa21, grz02}, which is known for
transferring long-range interactions to short-range ones, are also
performed to further justify our truncation. Within statistical
errors, the results for the global bond orientational order
parameter $\Psi_6$ and susceptibility $\chi_6$ obtained with
different truncating distances and the Ewald Summation are in good
agreement with each other. Relevant data with $L=64$ are compiled in
table \ref{t1} for comparison. Additional simulations with $L=128$,
and measurements of the pair distribution function $g(r)$ also
confirm the reliability of the truncation.

In this paper, most simulations are performed with molecular dynamics.
All results are obtained at a constant temperature with the $NVT$
ensemble based on the Nos\'{e}-Hoover Chain thermostat \cite{nos84,hoo85}.
The equation of the motion is solved via the five-point Nordsieck-Gear
predictor-corrector method. The time step $\Delta t$
in all the simulations is set to $0.01$. A shift of the conserved total
energy is within $0.0001\%$.

The initial configurations in our simulations consist of
particles uniformly distributed over the system box on a
triangular lattice. Before collecting data for the
measurements of physical observables, the system is carefully
equilibrated, especially in the critical regime. We monitor the
global bond orientational order parameter, and begin our measurement
after this order parameter reaches a steady value. For the
larger system ($N=16384$), for example, it takes $5\times 10^{5}$
time steps to thermalize the system. Only the configurations in
equilibrium are used for the measurements of observables,
extending over $9\times10^{6}$ time steps. In order to obtain independent configurations, the autocorrelation
function of the global bond orientational order parameter is
measured, and the correlation time is estimated to be $\tau\approx 2400$
time steps in the critical regime. Then the measurement is performed
every $2500$ time steps.

\begin{figure}[h]
\epsfysize=5.8cm \epsfclipoff \fboxsep=0pt
\setlength{\unitlength}{1cm}
\begin{picture}(10,7.0)(0,0)
\put(-4.3,0.5){\epsffile{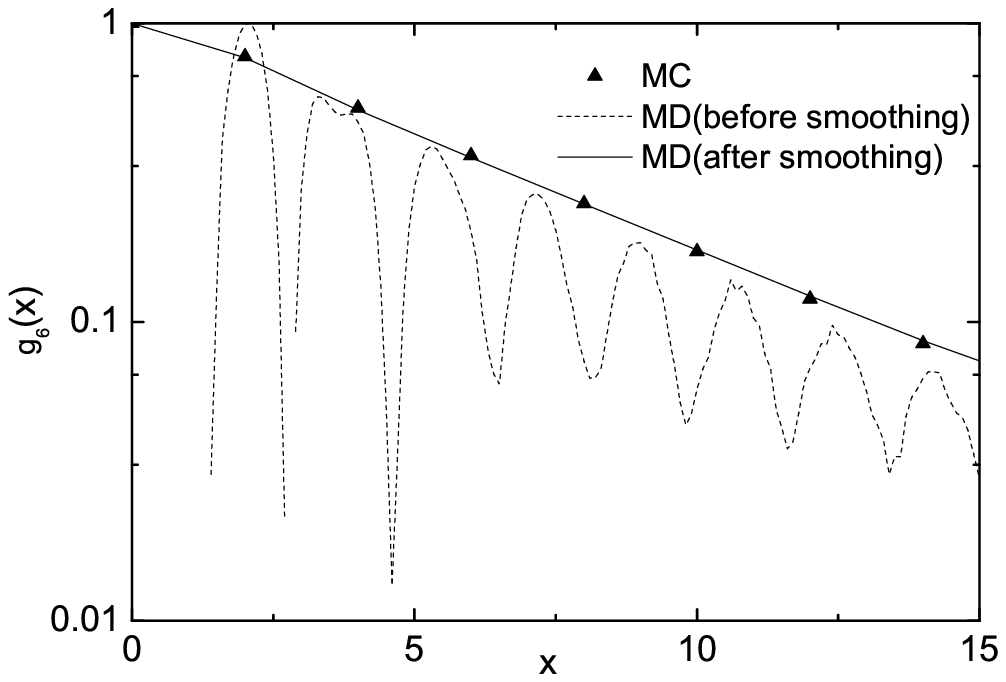}} \epsfysize=6.0cm
\put(4.5,0.3){\epsffile{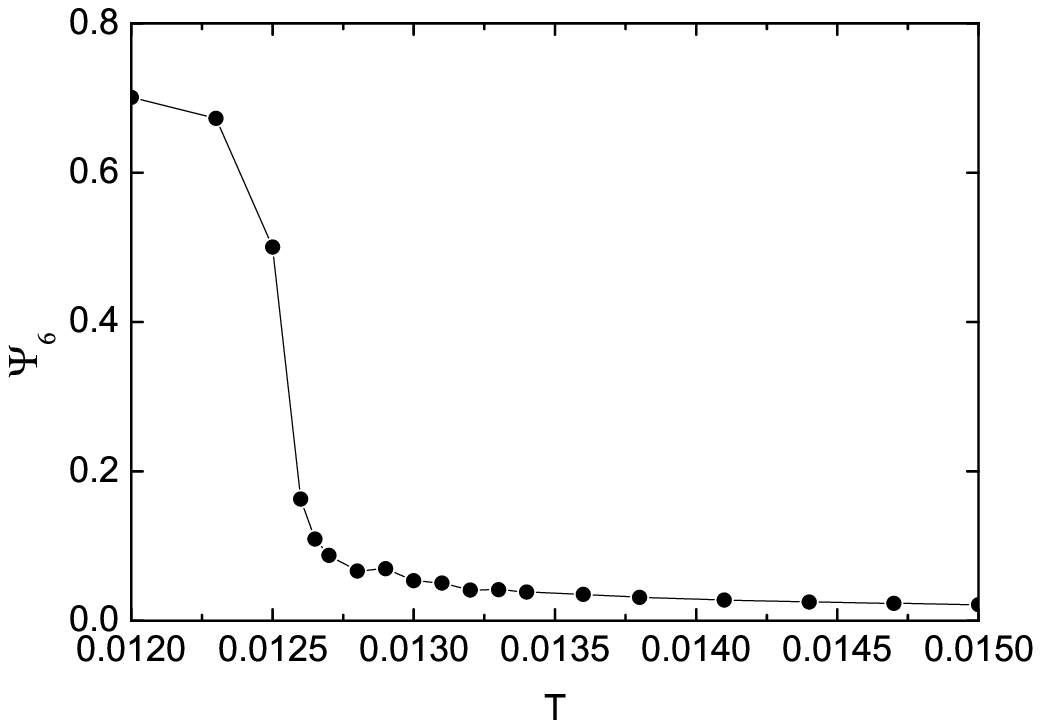}} \put(-2.5,0.2){(a)}
\put(6.2,0.2){(b)}
\end{picture}
\caption{(a) $g_{6}(x)$ obtained with molecular dynamics (MD) and
Monte Carlo (MC) simulations at $T=0.0150$ plotted vs. $x$ on a
semi-log scale. The smoothed curve is shifted upward for clarity.
The smoothing technique is described in Sec. II. (b) $\Psi_{6}$
plotted vs. $T$ on a linear plot. The bond orientational order
parameter increases abruptly around $T=0.0125$. The line fitted to
the circles is a guide to the eyes.}\label{f1}
\end{figure}

In order to confirm our molecular dynamics simulations, standard
Monte Carlo simulations are additionally performed. For example,
The bond orientational correlation functions from both molecular
dynamics simulations and Monte Carlo simulations are shown in
Fig.~\ref{f1}~(a). Both methods provide consistent curves, and it
shows that our molecular dynamics simulations indeed generate
proper ensemble distributions. Furthermore, dynamic Monte Carlo
simulations are carried out to extract the critical exponents for
the positional order parameter at the dislocation unbinding
temperature $T_{m}$.

\subsection{Observables}

The bond orientational symmetry of a solid can be described by the
six-fold global bond orientational order parameter $\Psi_{6}$ defined as
\begin{equation}\label{gbond}
\Psi_{6}=\langle|\frac{1}{N}\sum_{k=1}^{N}\psi_{6,k}|\rangle,
\end{equation}
where $N$ is the total number of the particles, $\langle
\cdot\cdot\cdot\rangle$ denotes the ensemble average or the time average in molecular dynamics simulations and Monte Carlo simulations, and the $\psi_{6,k}$ is the local bond orientational order parameter defined as
\begin{equation}\label{lbond}
\psi_{6,k}=\frac{1}{N_{k}}\sum_{j}\exp(i6\theta_{kj}).
\end{equation}
Here the sum j is over the neighbors of the particle k, and
$\theta_{kj}$  is the angle between $\overrightarrow{r_{kj}}$ (the
relative position vector of the particle k and j) and an arbitrarily
fixed reference axis. Neighbors are obtained with the Voronoi
polygon \cite {fra90}. The susceptibility of the bond orientational
order is defined as
\begin{equation}\label{sus}
\chi_{6}=N\langle\Psi_{6}^{2}\rangle.
\end{equation}

The hexatic phase is characterized by an algebraic decay of the
bond orientational correlation function defined as
\begin{equation}\label{g6x}
g_{6}(\overrightarrow{r_{1}}-\overrightarrow{r_{2}})
=\langle\psi_{6,k}^{*}(\overrightarrow{r_{1}})\psi_{6,k}(\overrightarrow{r_{2}})\rangle.
\end{equation}
In order to obtain an accurate value of the bond correlation length,
we smooth the bond orientational correlation function following Ref.
\cite{jas99}. We divide the volume of the system into stripes
with a width of $\Delta x $ and measure the
bond orientational correlation between different stripes.
\begin{equation}\label{smooth}
g_{6}(x)=\langle (\frac{1}{N(x)}\int_{0}^{L}{\mathrm
d}y'\int_{x-\Delta x/2}^{x+\Delta x/2} {\mathrm
d}x'\psi_{6,k}(\overrightarrow{r'}))^{*}\times(\frac{1}{N(0)}\int_{0}^{L}
{\mathrm d}y'\int_{-\Delta x/2}^{\Delta x/2} {\mathrm
d}x'\psi_{6,k}(\overrightarrow{r'}))\rangle ,
\end{equation}
where
\begin{equation}
N(x)=\int_{0}^{L}{\mathrm d}y'\int_{x-\Delta x/2}^{x+\Delta
x/2}{\mathrm d}x'\rho(\overrightarrow{r'}) ,
\end{equation}
\begin{equation}
\rho(\overrightarrow{r})=\sum_{i=1}^{N}\delta(\overrightarrow{r}-\overrightarrow{r_{i}}),
\end{equation}
and $L$ is the linear size of the system in the $y$ direction.
The temporal bond orientational correlation function characterizes
the time correlation of the bond orientational order parameter, and is
defined as
\begin{equation}\label{g6t}
g_{6}(t)=\langle\psi_{6,k}^{*}(t_{0})\psi_{6,k}(t_{0}+t)\rangle ,
\end{equation}
where $\psi_{6,k}(t_{0})$ and  $\psi_{6,k}(t_{0}+t)$ are the local
bond orientational order parameters measure at the time $t_{0}$ and
$t_{0}+t$ respectively, and the average is over $t_{0}$ in
equilibrium. In the hexatic phase, $g_{6}(t)$ also decays by a power
law \cite {zah00}.

The positional symmetry of solid can be described by a positional
order parameter defined as \begin{equation}\label{gpos}
\Psi_{pos}=\langle\frac{1}{N}\sum_{j=1}^{N}\exp(i\overrightarrow{G}\cdot\overrightarrow{r_j})\rangle,
\end{equation}
where $G$  is a reciprocal-lattice vector which gives
the first Bragg peak. In practice, we average the order parameter
over the six reciprocal vectors which correspond to the six vectors
connecting the six neighbors from the lattice site $j$.
The positional correlation function is defined as
\begin{equation}\label{poscor} g_{G}(|\overrightarrow{r}-\overrightarrow{r'}|)=\langle \exp(i\overrightarrow{G}\cdot(\overrightarrow{r}-\overrightarrow{r'}))\rangle.
\end{equation} In the hexatic phase, the positional correlation function decays exponentially. Finally, the pair distribution function is defined as \begin{equation}\label{gr}
g(r)=\frac{V}{N^{2}}\sum_{i,j\neq
i}\delta(\overrightarrow{r}-\overrightarrow{r_{ij}}) ,
\end{equation} where $V$ is the volume of the system.

\section{Computer simulation}

In this article, we perform extensive simulations of
two-dimensional melting in the $NVT$ ensemble with system sizes up to
$32786$ atoms, and find a strong evidence for the existence of
the hexatic phase in the dipole-dipole interacting system.
We first measure the spatial bond orientational correlation function
and susceptibility in the isotropic liquid phase and
compare the results with the predictions of the KTHNY
theory. This gives us estimates of the isotropic-anisotropic
transition temperature $T_i$. With this critical temperature in hand,
we further scan the parameter space, and observe an algebraic decay
of the spatial bond orientational correlation.
We also measure the temporal bond orientational correlation function,
and its behavior is in good agreement with the KTHNY theory.
In order to rule out a possible
coexistence phase and the finite-size effect, we perform a homogeneous
test and finite-size scaling analysis of the bond orientational order
parameter. The result is compatible with previous measurements. At
last, with Monte Carlo methods we simulate the short-time dynamics
of the positional order and estimate the exponent $\eta_{m}$,
and the value is also in agreement with the theoretical prediction.
All our results are compatible with the experiments and KTHNY theory,
and the hexatic phase is explicitly observed.

\subsection{Bond orientational order}

The bond orientational order parameter $\Psi_{6}$ offers a direct
description of the bond orientational order \cite{web95}. Assuming
$T_{i}$ is the transition temperature of the bond orientational
order and $T_{m}$ is the transition temperature of the positional
order, the bond orientational order parameter should vanish for
$T>T_{i}$. and take a finite value less than 1 for $T<T_{m}$. The
behavior of $\Psi_{6}$ at the temperatures between $T_{i}$ and
$T_{m}$ depends on the underlying melting scenario. If the
transition at $T_{i}$ is of first order, $\Psi_{6}$ increases
linearly from $T_{i}$ to $T_{m}$. If the melting scenario is of
KTHNY, i.e., the transition at $T_{i}$  is a Kosterlitz-Thouless
phase transtion, $\Psi_{6}$ then vanishes throughout the hexatic
phase for there doesn't exist a true long-range bond orientational
order. However, the finite-size effect in the simulations blurs this
distinction and prevents us drawing a clear conclusion.
Nevertheless, the measurement of the bond orientational order
parameter $\Psi_{6}$ does give us an estimated value of
$T_{i}\approx0.01250$. In the Fig.~\ref{f1} (b), $\Psi_{6}$ versus
$T$ is shown.

\begin{figure}[h]
\epsfysize=6.0cm \epsfclipoff \fboxsep=0pt
\setlength{\unitlength}{1cm}
\begin{picture}(10,7.0)(0,0)
\put(-4.3,0.5){\epsffile{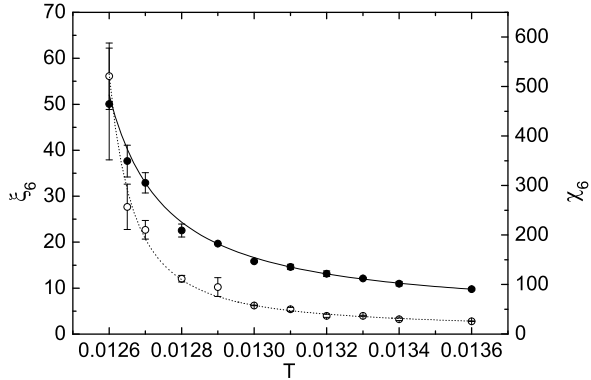}} \epsfysize=6.0cm
\end{picture}
\caption{Bond orientational correlation length (full symbols) and
susceptibility (open symbols) as a function of temperature. The curves show
the best fits of Eqs. (\ref {3a-1}) and (\ref {3a-2}) according to the
KTHNY theory. The fitted transition temperatures are $T_i=0.01237(16)$
and $ 0.01243(4)$
for the correlation length and susceptibility respectively} \label{f2}
\end{figure}

To further understand the phase transition at $T_{i}$,
we measure the correlation length and
susceptibility of the bond orientational order parameter in the
isotropic liquid phase for different temperature $T$. For the
measurements are carried out in the isotropic liquid phase, the
spatial bond orientational correlation function is independent of
the spatial directions. We extract the correlation length $\xi$
from the exponential decay of the bond orientational correlation
function smoothed with the technique described in Eq. (\ref{smooth}),
\begin{equation}
g_{6}(x) \sim \exp(-x/\xi).
\end{equation}
Subsequently, we compare our results with the predictions of the
KTHNY theory, i.e., an exponential singularity of the correlation
length and susceptibility,
\begin{equation}
\xi_{6}(\tau)=a_{\xi}\exp(b_{\xi}\tau^{-1/2}), \label {3a-1}
\end{equation}
\begin{equation}
\chi_{6}(\tau)=a_{\chi}\exp(b_{\chi}\tau^{-1/2}) , \label {3a-2}
\end{equation}
as $\tau=T-T_{i}\rightarrow0^{+}$. In Fig.~\ref{f2}, the
numerical data are fitted to the above exponential forms. The best
fit of the correlation length and susceptibility gives
$T_{i}=0.01237(16)$ and $T_{i}=0.01243(4)$ respectively.
These two values are in agreement with each
other within statistical errors, and are also consistent
with the previous estimated value
of $T_{i}$ from the global bond orientational order parameter.
Therefore, our results support the KTHNY prediction, even though
the statistical errors of the correlation length and
susceptibility are relatively large.

\subsection{The hexatic phase}

\begin{figure}[h]
\epsfysize=6.3cm \epsfclipoff \fboxsep=0pt
\setlength{\unitlength}{1cm}
\begin{picture}(10,7.0)(0,0)
\put(-4.3,0.5){\epsffile{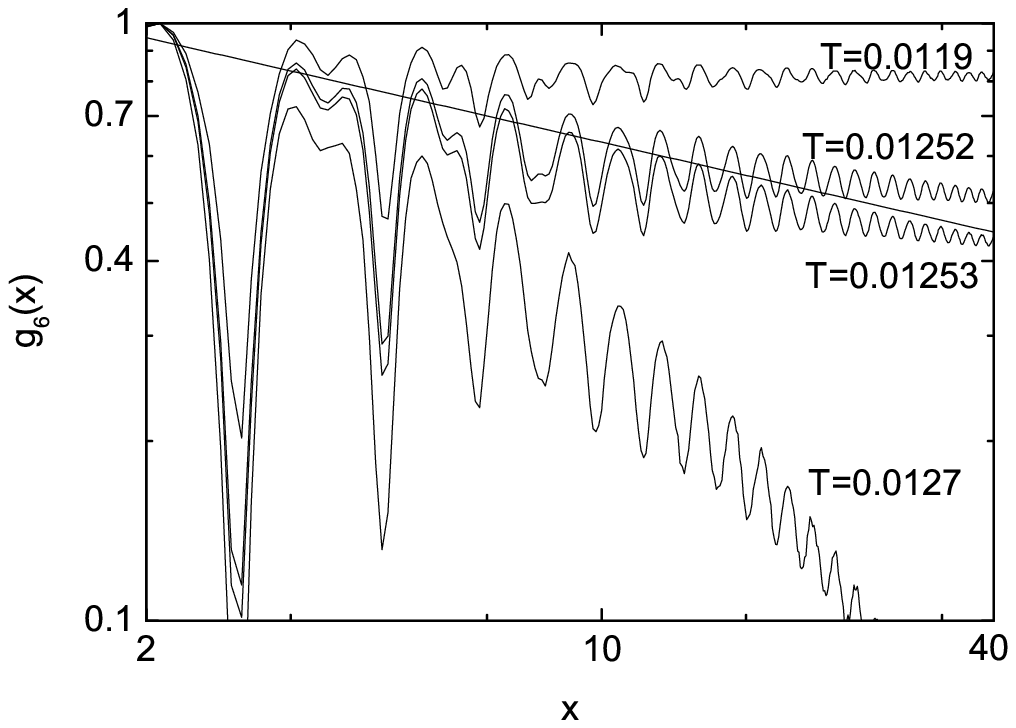}} \epsfysize=5.0cm
\put(4.5,1){\epsffile{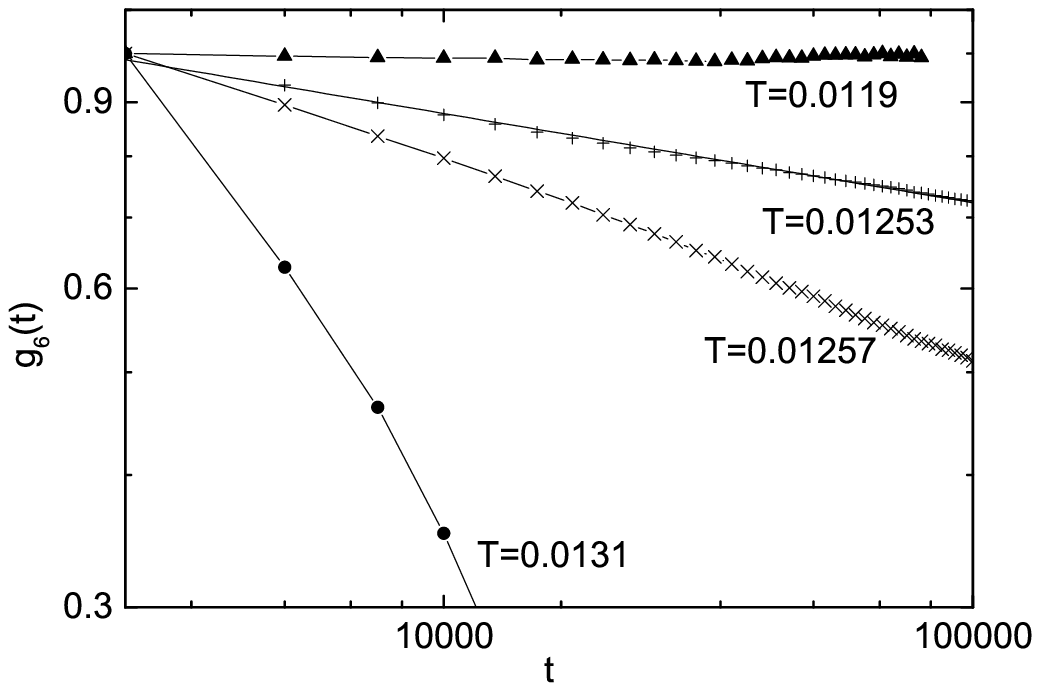}} \put(-2.5,0.5){(a)}
\put(6.2,0.5){(b)}
\end{picture}
\caption{(a) The spatial bond orientational correlation function
$g_{6}(x)$ plotted vs. $x$ on a double decimal log scale. The temperature
$T=0.0119$ is just before melting, $T=0.0127$ is typically
in the liquid phase, and $T=0.01252$ and $0.01253$ are in the
hexatic phase. The straight line with a slope of $-1/4$
is a guide to the eyes. (b) The temporal bond orientational
correlation function $g_{6}(t)$ plotted vs. $t$ on a
double-log scale. The temperature $T=0.0119$ is just before melting,
$T=0.0131$ is typically in the liquid phase, and $T=0.01253$ is in the
hexatic phase. $g_{6}(t)$ at another $T=0.01257$, which is slight
above the estimated $T_{i}$, is also shown. Lines fitted to the data
are to guide the eyes.}\label{f3}
\end{figure}

According to the KTHNY theory, the hexatic phase is characterized by
an algebraic decay of the bond orientational correlation function and
an exponential decay of the positional correlation function.
Therefore, we scan the parameter
space between $T_i$ and $T_m$, and measure the bond orientational
correlation function and positional correlation function. The bond
orientational correlation
function  is shown in Fig.~\ref{f3} (a). A clear evidence for the
existence of the hexatic phase is observed.

i) In the solid phase ($T=0.0119$), the correlation function rapidly
stabilizes to a constant, and it indicates that there is a true
long-range order of the bond orientational symmetry.

ii) In the hexatic phase ($T=0.01252$ and $0.01253$), the correlation
function shows an algebraic decay,
\begin{equation}
g_{6}(x)\sim x^{-\eta_6},
\end{equation}
and it indicates that there is a quasi-long-range
order of the bond orientational symmetry.

iii) In the liquid phase ($T=0.0127$), the correlation function decays
exponentially, and it indicates an isotropic state.

After smoothing the correlation function at $T=0.01253$ , we obtain
an exponent $\eta_{6}=0.252(6)$ from the slope of the curve,
and it is close to the value $\eta_{6}=0.25$ at the transition
temperature $T_i$ predicted by the KTHNY theory. If we assume $T=0.01253$ is just the transition temperature $T_i$, it is quantitatively in agreement
with the previous measurements in the preceding subsection. Nevertheless,
to obtain a more accurate value of $T_i$ , we need to consider more
carefully the finite-size effect. In the next subsection, we will locate the
transition temperature from the finite-size scaling.

In Fig.~\ref{f4}, the positional
correlation function and pair distribution function are shown
respectively. One may observe two different behaviors.

i) In the solid phase ($T_i=0.0119$), the positional correlation
function shows an algebraic decay, indicating a quasi-long-range
positional order in a two-dimensional solid, while the oscillation
in the pair distribution function persists over the entire range.

ii) In the hexatic phase ($T_i=0.01253$), the positional correlation
function decays quickly to zero, indicating that there exists no
positional order in the hexatic phase, and the oscillation in the pair
distribution function dies
off rapidly. The behaviors of the positional correlation function and
pair distribution function in the liquid phase are qualitatively the
same as in the hexatic phase.

The recent experiment reported in Ref. \cite {zah00} shows that the
dynamic behavior is also very important
in understanding the melting mechanism in two dimensions.
According to the KTHNY theory, in the solid phase
the temporal bond orientation correlation function
will rapidly stabilize to a constant, in the hexatic phase
it shows an algebraic decay with an exponent equal to
$\eta_{6}/2$ ,
\begin{equation}
g_{6}(t)\sim t^{-\eta_6/2},
\end{equation}
and in the liquid phase the temporal correlation function decays
exponentially \cite{nel83}. Such a behavior is indeed observed
in experiments, and it provides a strong evidence for the existence
of the hexatic phase. To our knowledge, such measurements have not been
performed in numerical simulations.

In order to deepen our understanding of two-dimensional melting
and further confirm our observation in numerical simulations of
static properties, the temporal bond orientation correlation function is
measured in our molecular dynamics simulations. The result is
shown in Fig.~\ref{f3}(b). Obviously, as the temperature changes
from $T=0.0119$ to $0.01253$, then to $0.0131$, the temporal bond
orientation correlation function follows the prediction of the
KTHNY theory, and are well consistent with the experimental
observation \cite {zah00}. The exponent measured from the slope of
the curve at $T=0.01253$ is $0.0843$, somewhat smaller than
the theoretical prediction $0.125$ at $T_i$. This probably suggests
that the anisotropic-isotropic transition temperature $T_{i}$
should be still slightly above the value $0.01253$, and our
measurements of the spatial and temporal bond orientational correlation functions may still carry certain finite-size effects.

\subsection{Finite-size scaling analysis}

Our measurements of the spatial and temporal bond correlation
functions provide us an explicit evidence for the existence of the
hexatic phase in two-dimensional melting. However, it is difficult
to extract an accurate transition temperature $T_i$ from
the correlations functions. One may directly measure the
correlation length in the isotropic liquid phase and then fit the
data to the ansatz in Eq. (\ref {3a-1}) and obtain the transition
temperature $T_i$. But this approach also suffers from the
difficulty that one can not obtain the correlation length at the
temperatures very close to $T_i$ \cite {gup92,zhe99b}. Meanwhile, due
to the finite-size effect, distinguishing between an algebraic and
an exponential decay might be problematic if the correlation
length is finite but much larger than the system size. Therefore,
to extract a more accurate disclination unbinding temperature
$T_i$ and to confirm the previous observation of the hexatic
phase, we perform a finite-size scaling analysis of the bond
orientational order parameter.

From the finite-size scaling form, the second moment
of the bond orientational order parameter can be written as
\begin{equation}
\langle\Psi_{6}^{2}\rangle\sim L^{-\eta_{6}}f(L/\xi_{6}) ,
\end{equation}
where $L=2\sqrt N$ is the linear size of the system and $\xi_{6}$ is
the bond correlation length. Since $\xi_{6}$ is divergent in the
hexatic phase, $\langle\Psi_{6}^{2}\rangle$ thus shows a power-law
behavior with respect to $L$ in the hexatic phase. In the liquid
phase, the power-law behavior will be modified by the scaling
function $f(L/\xi_{6})$.

\begin{figure}[h]
\epsfysize=6.0cm \epsfclipoff \fboxsep=0pt
\setlength{\unitlength}{1cm}
\begin{picture}(10,7.0)(0,0)
\put(-4.0,0.0){\epsffile{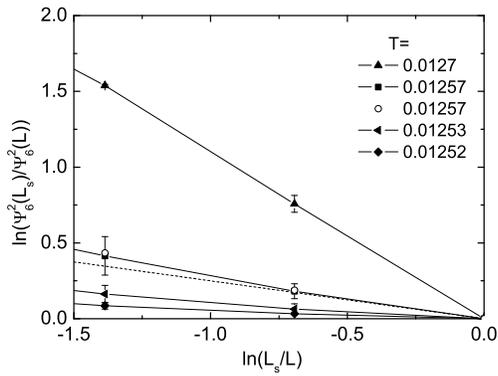}} \epsfysize=12.0cm
\end{picture}
\caption{The finite size scaling analysis of $\Psi_{6}^2$. $L=256$
or $512$ is the bulk linear size and $L_s$ is the size of the
subsystem. The dotted line with a slope of $-1/4$ is a guide to the
eyes. Open circles are the results from independent simulations with
periodic boundary conditions at $L=64,128,256$.} \label{f5}
\end{figure}

We measure the second moment of global bond orientational order
parameter with system size $L=64,128,256$ at $T=0.01257$ and perform
finite-size scaling analysis mentioned above. The open circles shown
in Fig.~\ref{f5} are the results. In order to save computation time,
we use the subsystem method introduced by the authors of Refs.
\cite{web95,bag96}. Here, a brief comment about the above
non-standard finite-size scaling analysis is needed. In principle,
the subsystem procedure still carries a second-order finite-size
effect induced by the finite bulk system size $L$. But this
second-order finite-size effect is negligibly small in practical
simulations \cite{web95}, and the procedure has been proved to be
reliable and may reduce computer times \cite{bag96}. We also carried
out the finite-size scaling analysis using subsystem method at
$T=0.01257$ to further justify this procedure, the result is shown
in Fig.~\ref{f5}. It is easy to observe that within statistical
errors, the data with periodic boundary conditions and from
subsystems are well consistent. With the subsystem method, we
measure $\langle\Psi_{6}^{2}\rangle$ at different temperatures with
a bulk linear size $L=256$ or $512$, and a total number of
particles ranging from $N=16384$ to $32768$. Then the system is
divided into small subsystems with a linear size $L_S$ \cite{web95}
and the bond orientational order parameter of each subsystem is
measured. The result is shown in Fig.~\ref{f5}.

To locate $T_{i}$, we assume $\eta_{6}=1/4$. In other words, we
search for a temperature which yields $\eta_{6}=1/4$, and then
assign this temperature to be $T_{i}$. The requirement of
$\eta_{6}=1/4$ yields the upper limit of $T_{i}$ \cite{nel83}.
Combining the results obtained in the preceding subsections, we
conclude $0.01253<T_{i}<0.01257$. To compare our results with
those in literatures, we convert $T_{i}$ to the dimensionless
parameter $\Gamma_{i}$, and obtain $68.707<\Gamma_{i}<68.927$. It
improves the values $T_{i}=62\pm3$ with a small system $N=256$
\cite{kal81} and $T_{i}=67.750$ with a relatively larger system
$N=961$ \cite{schun}. In Refs.\cite {kal81,schun}, the phase
transition is supposed to be of first order, and the values of
$T_{i}$ are obtained from the hysteresis in the temperature
dependence of energy, the existence of latent heat and the
thermodynamic nucleation of the solid from the supercooled liquid.
Our estimate of $T_{i}$ is based on the KTHNY theory, and is
much less affected by the finite-size effect.

\subsection{Ruling out the coexistence phase}

In principle, the $NVT$ molecular dynamics simulation can not obviate
the coexistence phase, and the superposition of the solid and liquid
phases may also produce the hexatic-like behavior. In order to exclude
this possibility, we apply the homogeneous test. We divide the system
into small subsystems and compute the susceptibility $\chi_{6}$ for all
subsystems \cite{str84}. If the system exhibits an inhomogeneous two-phase
coexistence, the probability distribution of $\chi_{6}$ at a
sufficiently small length scale could be modeled by a curve with two peaks,
which reflects a combination of solid and fluid distributions.
On the other hand, if the system is homogeneous, varying the size
of the subsystems should not lead to any qualitative change
in the probability distribution of $\chi_{6}$, i.e.,
the curve should always remain with a single peak.

\begin{figure}[h]
\epsfysize=6.0cm \epsfclipoff \fboxsep=0pt
\setlength{\unitlength}{1cm}
\begin{picture}(10,7.0)(0,0)
\put(-4.0,0.5){\epsffile{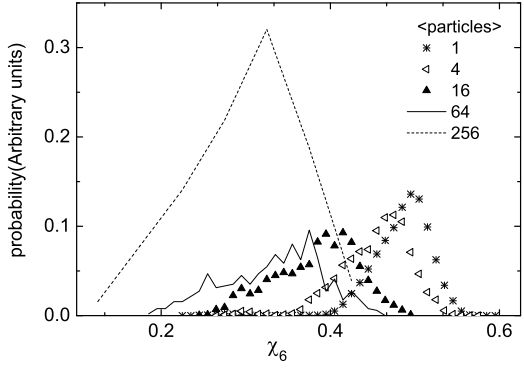}} \epsfysize=6.0cm
\end{picture}
\caption{The probability distribution of $\chi_{6}$
in the hexatic phase at $T=0.01252$. The symbols
in the figure indicate the mean numbers of particles in
different subsystem. } \label{f6}
\end{figure}

We have measured the possibility distribution of $\chi_{6}$
in the hexatic phase at $T=0.01257, 0.01253, 0.01252$ ,
and in order to compare the result with that in the homogeneous phase,
we also perform a simulation at an extra temperature
$T=0.0100$ corresponding to the cool solid phase. No
qualitatively change is found at these temperatures. This test
rules out the existence of a coexistence phase, and confirms
the observation of the hexatic phase in the previous subsections.
The result at $T=0.01252$ is shown in Fig.~\ref{f6}. It is clearly
seen that varying the system size doesn't change
the shape of distributions, but only shifts the peak of
the probability distribution.

\subsection{Dynamic Monte Carlo simulations}

In the last decade, it has been discovered that already in a
macroscopic short-time regime emerges the universal scaling
behavior \cite{zhe98,luo98b,jas99a,zhe99,wat04}. Measurements now are
carried out at the early stage of the time evolution, therefore
one does not suffer from critical showing down. The dynamic
scaling form of the second moment of the positional order
parameter below the dislocation unbinding transition temperature
$T_{m}$ is
\begin{equation}
\Psi_{pos}^{2}(t,L)=b^{-\eta_{m}}\Psi_{pos}^{2}(b^{-z}t,b^{-1}L) ,
\end{equation}
where $t$ is the evolution time, $z$ is the dynamic critical exponent,
and
$b$ is an arbitrary rescaling factor. For a sufficient large
$L$, this dynamic scaling form is reduced to
\begin{equation}\label {Psi}
\Psi_{pos}^{2}(t)\sim t^{-\eta_{m}/z} .
\end{equation}
From a finite-size scaling analysis of the time-dependent Binder
cumulant
\cite {zhe98}
\begin{equation}
U_{pos}(t)=\frac{\Psi_{pos}^{4}}{(\Psi_{pos}^{2})^{2}}-1\ ,
\end{equation}
one obtains
\begin{equation}\label{cumulant}
U_{pos}(t)\sim t^{d/z}/L^d,
\end{equation}
where $d$ is the dimension of the system. The dynamic critical exponent
$z$ can be estimated from Eq.(\ref{cumulant}), and with $z$ in hand, the
static exponent $\eta_{m}$ can be obtained from Eq. (\ref {Psi}).

Now, we turn to locate the transition temperature $T_m$.
As the temperature increases, $T_{m}$ is characterized by the dislocation
unbinding which breaks the quasi-long-range positional symmetry.
Therefore, one may measure the correlation function of the positional
order parameter to estimate $T_{m}$, for the positional correlation
become short-range at $T_{m}$. Nevertheless, this method
suffers from the difficulty that one needs to do simulations
in the critical region. Even for the hard disk model,
in which the thermodynamic quantities
are only determined by the density $\rho$ of disks, it is still
not easy to locate $\rho_{m}$ accurately. $\rho_{m} \approx 0.933$
is reported in Ref.\cite{jas99a}, while
$\rho_{m}=0.910(2)$ is given in Ref. \cite{wat04}.

Alternatively, a dynamic technique for locating $T_{m}$ is applied
in the experiments reported in Ref. \cite{zah00}. In our computer
simulations, we follow Ref. \cite{zah00} and adopt the dynamic
criterion for $T_{m}$, since the method is relatively simple, and
may provide direct comparison with experiments. We first introduce
the $2D$ Lindemann parameter \cite{bed85,zah99}
\begin{equation}\label{linde}
\gamma_{m}=\langle|\overrightarrow{u}(\overrightarrow{r}+\overrightarrow{a_{0}})-\overrightarrow{u}(\overrightarrow{r})|^{2}\rangle\times\pi
n ,
\end{equation}
where $\overrightarrow{a_{0}}$ is the lattice spacing vector,
$\overrightarrow{r}$ is the positional vector, $\overrightarrow{u}$
is the displacement field and  $n$ is the $2D$ volume fraction of
particles. Initially, the system is set on a perfect triangular
lattice. In numerical simulations, we gradually warm up the
system with the velocity rescaling procedure. At each $T$, the
system is equilibrated to the thermal equilibrium.
Then we measure the Lindemann parameter at different temperatures.
In general, a sharp growth of $\gamma_{m}$ indicates vanishing of
the positional symmetry. Such a Lindemann criterion in $3D$ systems
is a well-established and justified procedure for locating the
melting temperature $T_m$, although it was unclear in two dimensions
\cite{mer66}. In $1985$, Bedanov and Gadiyak improved the definition
of the Lindemann parameter to the form in Eq. (\ref{linde})
and demonstrated in the simulations of electron and dipole systems
that when $\gamma_{m}$ rises up to a ¡®critical¡¯ value $0.12$, the
melting takes place \cite{bed85}. At the melting point $T_{m}$,
which is more clearly identified by the sudden drop of the
positional correlation length, a sharp growth of $\gamma_{m}$ is
induced by the leap of disclination number and self-diffusion
constant. Therefore, this local quantity is relevant to the melting.
Recently, Zahn {\it et al.} applied this criterion to their
experiments \cite{zah99,zah00}, and the results are in agreement
with those from numerical simulations \cite{bed85}. The Lindemann
parameter may provide at least a first estimate of the melting
temperature $T_m$. Due to its efficiency and simplicity, the Lindemann
criterion has been applied to different systems for locating the melting
point \cite{sai03,tya04,ras05}.

\begin{figure}[h]
\epsfxsize=8.5cm \epsfysize=6.2cm \epsfclipoff \fboxsep=0pt
\setlength{\unitlength}{1cm}
\begin{picture}(10,1.5)(-0.5,0)
\put(-4.8,0.5){\epsffile{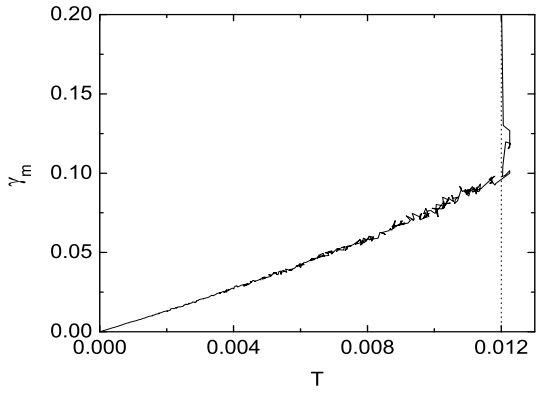}} \epsfysize=7.0cm
\end{picture}
\caption{The Lindemann parameter $\gamma_{m}$ vs. $T$. The critical
temperature $T_{m}=0.0120$ is visualized by the vertical dotted line.}
\label{lindemann}
\end{figure}

Here we locate $T_{m}$ with the Lindemann criterion. Four runs are
performed in order to estimate $T_m$ . One of them is shown in Fig.
~\ref{lindemann}. We estimate $T_{m}=0.0120(2)$. The hexatic phase
of the dipolar system lies in a range of the phase diagram, between
$0.01253<T_{i}<0.01257$ and $T_{m}=0.0120(2)$. This is comparable
with that of the hard disk model, $\rho_{i}=0.899(1)$ \cite{jas99},
and $\rho_{m}\approx 0.933$ in Ref. \cite{jas99a} while
$\rho_{m}=0.910(2)$ in Ref. \cite{wat04}. $T_{i}$ and $T_{m}$ may
overlap for small systems, and this is one reason why the hexatic
phase was not observed in some previous studies.

Now we perform dynamic Monte Carlo simulations at the transition
temperature $T_{m}$. The reason we perform Monte
Carlo simulations is that the dynamic scaling forms in Eqs. (\ref {Psi})
and (\ref {cumulant}) may not hold in the dynamic process of Nos\'{e}-Hoover chain molecular dynamics simulations. It seems that the Nos\'{e}-Hoover Chain method is originally devised for equilibrium simulations and contains techniques violating the dynamic scaling behavior. In comparison to this, the dynamic scaling behavior in Monte Carlo simulations has been extensively justified.

\begin{figure}[h]
\epsfysize=5.0cm \epsfclipoff \fboxsep=0pt
\setlength{\unitlength}{1cm}
\begin{picture}(8.5,7.5)(0,0)
\put(-4.3,0.8){\epsffile{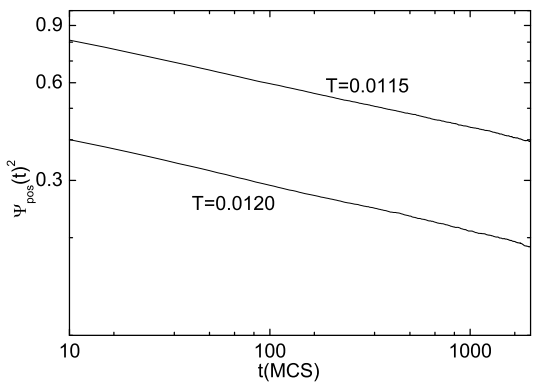}} \epsfysize=5.1cm
\put(4.5,0.8){\epsffile{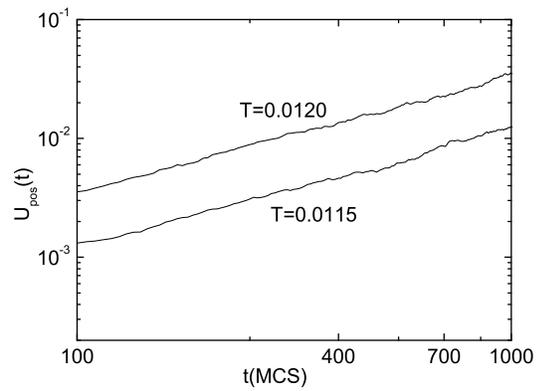}} \put(-3.5,0.2){(a)}
\put(5.5,0.2){(b)}
\end{picture}
\caption{(a) $\Psi_{pos}^{2}(t)$ plotted vs. $t$ on a
double-log scale. The lower curve is shifted downward for
clarity. (b) $U_{pos}(t)$ plotted vs. $t$ on a
double-log scale. The upper curve is shifted upward for
clarity.}\label{f7}
\end{figure}

In Monte Carlo simulations, the system initially at an ordered state is released to the dynamic evolution with the Metropolis algorithm, and then the time-dependent
$\Psi_{pos}^{2}$ and $U_{pos}$ are measured. By fitting
$\Psi_{pos}^{2}(t)$ and $U_{pos}(t)$ to Eqs. (\ref {Psi}) and (\ref
{cumulant}), both the dynamic exponent $z$ and static exponent
$\eta_{m}$ can be determined. For comparison, we also perform the
same simulations at another temperature $T=0.0115$. The results are
shown in Fig.~\ref{f7} (a) and (b).

From $U_{pos}(t)$ in Fig.~\ref{f7} (b), we estimate $z=1.910(70)$,
and from $\Psi_{pos}^{2}(t)$ in Fig.~\ref{f7} (a), we measure
 $\eta_{m}/z=0.143(5)$. Combining these results, we deduce
$\eta_{m}=0.273(20)$. This value is also in agreement with the
prediction ($1/4\leq\eta_{m}\leq1/3$ ) based on the KTHNY theory
\cite{str88}.

\section{Conclusion}

We present molecular dynamics and Monte Carlo simulations of
two-dimensional melting with dipole-dipole interactions. An
algebraic decay is observed for both the spatial and temporal bond
orientational correlation functions in an intermediate temperature
regime, and this serves as an explicit evidence for the existence of
the hexatic phase.

To obtain a relatively accurate disclination unbinding temperature
$T_{i}$,
we perform a finite-size scaling analysis for the bond
orientational order parameter. The result $0.01253<T_{i}<0.01257$
improves the value from a direct fit of the correlation length to
the exponential ansatz. In addition, by analyzing the probability
distribution of the bond orientational susceptibility $\chi_{6}$,
a possible coexistence phase is ruled out.

At last, from dynamic behavior of the Lindemann parameter, the
dislocation unbinding transition temperature is estimated to be
$T_{m}=0.0120(2)$. We also perform dynamic Monte Carlo simulations
of the positional order parameter and the time-dependent cumulant.
From the power-law behavior of these quantities, we determine the
exponents $\eta_{m}=0.273(20)$ and $z=1.910(70)$ .

In summary, a clear evidence for the existence of the hexatic
phase is observed for two-dimensional melting with dipole-dipole
interactions, and all the static and dynamic behaviors of the
system are compatible with recent experiments and the KTHNY
theory.

{\bf Acknowledgements:} The authors would like to thank A. Jaster
for helpful discussions and suggestions. Numerical Computations
have been performed at the Center for Engineering and Scientific
Computation, Zhejiang University. This work is supported in part
by NNSF and SRFDP (China) and DFG (Germany).


\begin{table}[h]\centering
\begin{tabular}{l  |  l | l | l | l}\hline\hline
 &   $\Psi_6 (T=0.0150)$  &  $\Psi_6 (T=0.0125)$ & $\chi_6 (T=0.0150)$ & $\chi_6 (T=0.0125)$ \\
 \hline
 $r_t=10$ & $0.0842(25)$ & $0.684(4)$ & $9.14(49)$ & $479(6)$  \\
\hline
 $r_t=20$& $0.0849(35)$& $0.680(4)$ & $9.34(38)$ & $475(5)$
\\
\hline Ewald Summation & $0.0859(37)$& $0.680(1)$ & $9.50(84)$ & $474(2)$ \\

\hline\hline
\end{tabular}
\caption{ The global bond orientational order parameter $\Psi_6$ and
susceptibility $\chi_6$ measured by truncating the potential at
$r_t=10$, $r_t=20$ and with the Ewald Summation to deal with the
potential. The linear size is $L=64$, and the temperature is
$T=0.0150$ and $T=0.0125$.} \label{t1}
\end{table}

\end{document}